\date{}
\documentstyle[prl,aps,preprint,epsf]{revtex}

\begin{document}
\draft
\title {Universality in quantum parametric correlations}
\author{\large P. Leb{\oe}uf \hspace{-0.2cm} $^a$ and M. Sieber
        \hspace{-0.2cm}  $^{a,b}$}
\address{a. Laboratoire de Physique Th\'eorique et Mod\`eles Statistiques 
  \footnote{Unit\'e de recherche de l'Universit\'e de Paris XI associ\'ee au 
  CNRS}, B\^at. 100, \\ 91405 Orsay Cedex, France \\
  b. Abteilung Theoretische Physik, Universit\"at Ulm, 89069 Ulm, Germany
  \footnote{New address: Max-Planck-Institut f\"ur Physik komplexer Systeme,
N\"othnitzer Str.\ 38, 01187 Dresden, Germany}}
\maketitle

\begin{abstract}
We investigate the universality of correlation functions of chaotic and
disordered quantum systems as an external parameter is varied. A new,
general scaling procedure is introduced which makes the theory invariant
under reparametrizations. Under certain general conditions we show that this
procedure is unique. The approach is illustrated with the particular case of
the distribution of eigenvalue curvatures. We also derive a
semiclassical formula for the non-universal scaling factor, and give an
explicit expression valid for arbitrary deformations of a billiard system.
\end{abstract}


\newpage

Several criticisms were formulated many years ago concerning the
applicability of random matrix theory (RMT) in the description of the
behavior of complex quantum systems such as the atomic nucleus. The basic
question was: should one trust the predictions of RMT at all if already the
average density of states (Wigner's semi-circle law) does not give a good
description of real systems? We now know that RMT describes the universal
behavior of local fluctuation properties. These have been shown to be, in
the limit of large dimensions, invariant for a large class of ensembles of
matrices, while the average spectral density is ensemble dependent and
therefore non-universal \cite{dyson,balian}. For a given physical system, it
is now well established that in order to eliminate the system dependent
features and observe universal fluctuations one should consider, instead
of the original eigenenergies $E_j$, $j = 1, 2, \ldots$, the unfolded
spectrum
\begin{equation} \label{eunfolding}
\epsilon_j = {\bar N} (E_j) \ ,
\end{equation}
where
\begin{equation} \label{step}
{\bar N} (E) = \int^{E} {\bar \rho} (E') d E'
\end{equation}
is the integrated average density of states \cite{bg}. By construction the new
energies $\epsilon_j$ have unit mean level spacing. If the system has a
classical analog with  $n$ degrees of freedom described by
the Hamiltonian $H ({\vec p},{\vec q})$, ${\vec p} = (p_1, \ldots, p_n)$, 
${\vec q} = (q_1, \ldots, q_n)$, then to leading order in a semiclassical
expansion the average density of states is given by the Thomas-Fermi
approximation
\begin{equation} \label{weyl} 
{\bar \rho} (E) \approx \frac{1}{(2 \pi \hbar)^n} \int\int d^n p \ d^n q \
\delta[E - H({\vec p},{\vec q}) ] \ .
\end{equation}
This expression allows for an explicit implementation of the unfolding
procedure (\ref{eunfolding})-(\ref{step}). 

More recently, extensions of the universal behavior of disordered as well as
chaotic quantum systems have been developed to include parametric
correlations and fluctuations of the energy levels as some real parameter
$X$ controlling the dynamics is varied (see section III-H of
\cite{gmw} for a recent review). These extensions are of physical importance
because many response functions such as the magnetic susceptibility or the
conductance may be expressed as parametric correlations. Exactly as for the
usual fluctuations computed at fixed parameter values, the mean properties
of the flow of energy levels $\epsilon_j (X)$ when $X$ is varied are
system-dependent and therefore non-universal. An appropriate scaling
procedure is again necessary in order to extract the universal behavior.

While the average velocity of the levels is set to zero if the energy
unfolding (\ref{eunfolding}) is implemented for arbitrary $X$, what now
characterizes the flow $\epsilon_j (X)$ is the typical slope of the
eigenvalues with respect to $X$. This typical slope is not only
system-dependent, but it is moreover parameter-dependent. In fact, the
curves $\epsilon_j (X)$ as well as their average properties vary when
expressed as a function of a different parameter $Y (X)$. Therefore, in
order to have universality in the theory it is necessary to show the
existence of a parametric representation of the elementary quantum objects
satisfying two basic conditions: (i) it scales out system-dependent features
and, (ii) it is invariant under reparametrizations. This particular
parametric representation should moreover be unique (it should be ``the''
parametrization), since the existence of several different invariant
parameters associated to $X$ would destroy the universality of the
correlators.

The way to construct such a parameter is as follows. Consider the variance
of the distribution of the parametric velocities computed at fixed $X$ in
some window $(\epsilon-\Delta \epsilon/2,\epsilon+\Delta \epsilon/2)$,
containing $N$ levels, in which the statistics are evaluated
\begin{equation}\label{variance}
\langle v_{X}^2 \rangle  = (1/N)
\sum_j \left( \partial \epsilon_j /\partial X \right)^2 \ .
\end{equation}
This function characterizes the non-universal mean properties of the flow
$\epsilon_j (X)$ in the window we are considering. More precisely, it
characterizes the response of the energy levels to an external perturbation
and, following Thouless \cite{thou}, may be interpreted as a generalized
conductance of
the system \cite{am,sa1}. Then, to scale the spectral flow and eliminate the
system-dependent characteristics we introduce the parameter
\begin{equation}\label{mu}
\mu  = \int_{X_0}^X \sqrt{\langle v_{X}^2 \rangle } \ dX
\end{equation}
where $X_0$ is some reference value. The definition of $\mu$ is
universal, independent of the nature of $X$. If the system is
reparametrized by introducing a new parameter $Y (X)$, then
$\sqrt{\langle v_{X}^2 \rangle } = \sqrt{\langle v_{Y}^2 \rangle } \ 
|\partial Y /\partial
X |$. The last factor $|\partial Y /\partial X |$ is compensated
by the Jacobian of the transformation in Eq.(\ref{mu}) and we have
\begin{equation} \label{univ}
\mu  = \int_{X_0}^X \sqrt{\langle v_{X}^2 \rangle } \ dX = 
       \int_{Y_0}^Y \sqrt{\langle v_{Y}^2 \rangle } \ dY 
\end{equation}
for any transformation $Y (X)$. Notice that the prescription (\ref{mu})
yields a parametrization which makes the velocity variance identical to one,
$\langle v_\mu^2 \rangle  \equiv 1$, at all parameter values $\mu$.

We investigate now the uniqueness of this parametrization. 
It is actually possible
to construct an infinite number of invariant parameters, according to
$$
\mu_m = {\cal N}_m \int_{X_0}^X \langle v_{X}^{2 m} \rangle ^{1/2 m} \ dX \ ,
$$
with $m=1,2,\ldots$ (all these parameters satisfy the analogue of
Eq.(\ref{univ})). The normalization constant ${\cal N}_m$ is defined in
terms of the $2 m$-th moment of a Gaussian distribution with variance one,
${\cal N}_m = [(2 m - 1)!!]^{-1/2 m}$. In general these parameters define
different functions of $X$. However, there is one particular case for which
they are all identical (and coincide with the simplest one $m=1$ of
Eq.(\ref{mu})), and this is when the distribution of velocities is Gaussian.
This distribution is expected to hold for generic fully chaotic systems. It
can be easily seen to apply to the following parametric random matrix model
\begin{equation}\label{hrmt}
H = \cos X \, H_1 + \sin X \, H_2 \ ,
\end{equation}
where $H_1$ and $H_2$ are two independent random matrices belonging to one
of the three universality classes $\beta=1$, $2$ or $4$ (orthogonal, unitary
and symplectic, respectively) \cite{aw}. Furthermore, a Gaussian
distribution of velocities also holds for weakly disordered metallic
systems, where it has been explicitly demonstrated \cite{sa1}.

We have therefore established the existence of a unique,
{\sl parametric-invariant scaling} procedure for the restricted class of
systems having a Gaussian distribution of velocities. Conversely, this
distribution of velocities becomes a necessary condition for universality.
In non-generic fully chaotic or disordered systems where the velocities are
not Gaussian distributed (like for example in strongly disordered electronic
systems or banded random matrix models \cite{fyod}) the different invariant
parametrizations are not equivalent, and the universality is lost.

All these considerations have their analog in the usual unfolding procedure
at fixed parameter values. The motivation to unfold the spectrum according to
the prescription (\ref{eunfolding})-(\ref{step}) -- and not according to the
more "primitive" one ${\bar \rho} (E_j) E_j$ -- is not only because it
fixes the average mean spacing to one (both prescriptions do), but more
basically because it makes the new spectrum $\epsilon_j$ invariant under
reparametrizations of the energy.

As an illustration let us consider two well known correlators, the velocity
correlation function
\begin{equation}\label{v-v}
c_\mu (\nu) = \left\langle \frac{\partial \epsilon_j}{\partial \mu}
(\mu_0) \frac{\partial \epsilon_j}{\partial \mu} 
(\mu_0 + \nu) \right\rangle_{\mu_0,j}  \ , 
\end{equation}
and the distribution $p (k_\mu)$ of curvatures
\begin{equation}\label{curvature1}
k_\mu = \frac{1}{\pi \beta} \frac{\partial^2 \epsilon_j}{\partial
\mu^2} \ . 
\end{equation}
These quantities were investigated in the past for fully chaotic systems and
in random matrix theory, as well as in disordered systems
\cite{grmn,am,sza,sa1,zd,been,oppen,lr,blm}. The parametric
correlators were computed in terms of the rescaled parameter
\begin{equation}\label{kappa}
x = \sqrt{\langle v_{X}^2 \rangle } \ X \ ,
\end{equation}
which in general is not invariant under reparametrizations and may produce
non-universal results. For example, curvatures with respect to the parameter
$\mu$ and $x$ 
($k_x = (\partial^2 \epsilon_j /\partial X^2)/(\pi \beta \langle v_{X}^2 
\rangle )$)
are related by
\begin{equation} \label{prac}
k_\mu = k_x - \frac{(\partial \epsilon_j /\partial X)}{2 \pi \beta}
\frac{\left( \partial \langle v_{X}^2  \rangle  /\partial X \right)}
{\langle v_{X}^2  \rangle ^2} \ .
\end{equation}
Thus, what is expected to be universal is not the distribution of $k_x$ but
the distribution of the particular combination given on the r.h.s. of
Eq.(\ref{prac}). The lack of reparametrization invariance of $k_x$ was
properly emphasized and nicely illustrated in Ref.\cite{lr}.

It follows from the definition (\ref{kappa}) that the parameter $x$
coincides with $\mu$ if the function $\langle v_{X}^2 \rangle $ is
stationary with respect
to $X$, i.e. it is independent of $X$. Computations done on a stationary
spectrum having a Gaussian distribution of velocities using the parameter
$x$ are therefore correct in the sense that the results obtained are
expected to be universal. Because the Hamiltonian (\ref{hrmt}) satisfies
this property, the distribution
\begin{equation}\label{crmt}
 p(k) = {\cal N}_\beta (1 + k^2)^{-(\beta + 2)/2} 
\end{equation}
obtained from that model in Ref.\cite{oppen} {\sl is} the universal
distribution for the curvature (here ${\cal N}_\beta$ is a normalization
constant). On the other hand, in the generic situation when parametric
correlations are computed in a system where $\langle v_{X}^2 \rangle $
varies with $X$
(non-stationary spectrum), the use of the rescaling (\ref{kappa}) leads to
non-universal results and the observed distribution changes with $X$ exactly
as observed in Ref.\cite{lr}, unless the correct parametric-invariant
scaling (\ref{mu}) is used. To illustrate this point we show in Fig.1 the
curvature distribution of $k_\mu$ and compare it with that of $k_x$ for the
Robnik-lima\c{c}on billiard (the ``worst case'' found in Ref.\cite{lr} is
considered). The use of the universal parameter $\mu$ produces a dramatic
change on the distribution. We believe that the agreement with
Eq.(\ref{crmt}) will be further improved by going higher in the spectrum.

The functional values of statistics involving only first derivatives with
respect to the parameter, like in Eq.(\ref{v-v}), are invariant under
reparametrizations using the parameter $x$, and one would then believe that
both scalings are equivalent in this particular case. However, if
$\langle v_{X}^2 \rangle $ varies with $X$ the function looks different
when plotted
against $x$ or $\mu$, because $x (X)$ is different from $\mu (X)$. The use
of the appropriate parametric-invariant scaling (\ref{mu}) is therefore
necessary even for correlators involving first derivatives only.

As for the density of states Eq.(\ref{weyl}), it would be desirable to have
an explicit expression allowing for a direct computation of the
non-universal function $\langle v_{X}^2 \rangle $ (the generalized
conductance). Such an expression may be obtained from semiclassical
estimates of off-diagonal matrix
elements by applying results of RMT \cite{aw} or by comparing semiclassical
computations of the parametric density correlation function with results
obtained in disordered metallic systems \cite{sa1,bk}. It is however
possible to derive it from a direct semiclassical calculation based on the
Gutzwiller trace formula. The starting point is the counting function for
the unfolded spectrum 
$N (\epsilon,X) = \sum_j \Theta (\epsilon - \epsilon_j (X))$ 
($\Theta$ is the Heaviside function). We define the velocity density as
\begin{equation} \label{c2}
\rho_v^\eta (\epsilon) = - \frac{d N^\eta}{d X} = \frac{1}{\pi} \sum_j
\frac{\eta}{(\epsilon-\epsilon_j)^2 + \eta^2} \left( 
\frac{\partial \epsilon_j}{\partial
X} \right) , 
\end{equation}
where for convenience we have replaced the delta function by an
$\eta$-smoothed Lorentzian. From this we obtain
\begin{equation} \label{c4}
\sum_j \delta (\epsilon-\epsilon_j) \left( \partial \epsilon_j /\partial X
\right)^2 = 
\lim_{\eta \rightarrow 0} 2 \pi \eta \ [\rho_v^\eta (\epsilon)]^2 \ ,
\end{equation}
which is our starting point for the semiclassical calculations since the
average of the l.h.s. defines the average variance $\langle v_X^2 \rangle $.
To leading order in $\hbar$, $N^\eta (\epsilon,X)$ for a chaotic system is
given by \cite{gutz}
$$
N^\eta (\epsilon,X) = \epsilon + \frac{\hbar}{\rm i} \sum_p \frac{A_p}{T_p} 
{\rm e}^{{\rm i} S_p /\hbar - \eta t_p /\hbar} \ .
$$
The sum is over all the periodic orbits of the classical system, $A_p$ is an
amplitude which depends on their stability and $S_p$ and $T_p$ are their
action and period, respectively. Furthermore,
$t_p = |\partial S_p / \partial \epsilon| = |T_p / \bar{\rho}|$
and the sum runs over positive and negative values of $p$. Deriving this
expression with respect to $X$ according to Eq.(\ref{c2}), replacing in
Eq.(\ref{c4}), averaging in a small energy window and keeping only the
diagonal part of the sum one obtains a semiclassical
estimate for $\langle v_X^2 \rangle$. An analogous calculation was made in 
Ref.\cite{efkamm} to compute the variance of diagonal matrix elements of an 
arbitrary operator $A$. In our case $A=\partial H/\partial X$, and we obtain
\begin{equation}\label{c6}
\langle  [\rho_v^\eta (\epsilon)]^2  \rangle  = \frac{2}{h^2}
\int_0^\infty \frac{K_D}{T^2}
\langle Q_p^2 \rangle  {\rm e}^{-2 \eta T/(\bar{\rho} \hbar)} \ dT,
\end{equation}
with $K_D (T) = h^2 \langle \sum_p |A_p|^2 \delta (T-T_p) \rangle $, and
$\langle Q_p^2 \rangle = \sum_p Q_p^2 |A_p|^2 / \sum_p |A_p|^2$ with 
$Q_p = \partial S_p/\partial X|_\epsilon$. The sum
runs over orbits having a period $T_p$ between $T$ and $T+dT$. It can be
shown that the vanishing of the off-diagonal terms follows
from RMT, if we assume that the semiclassical theory can reproduce the
RMT results for parametric correlations. Furthermore, the variance of the
distribution of $Q_p$ is simply proportional to the period of the orbit
\cite{gsbswz}, $\langle Q_p^2 \rangle = \alpha T$.

The function $K_D (T)$ has also a linear dependence on the period 
\cite{berry}, 
$K_D (T) = 2 \ T/\beta$. From this and from Eqs.(\ref{c6}) and (\ref{c4}) we
get the final expression for the variance of the velocity \cite{efkamm}
\begin{equation} \label{cvariance}
\langle  v_{X}^2  \rangle  = \frac{\alpha \bar{\rho}}{\beta \pi \hbar} \ .
\end{equation}
It is moreover easy to see from the previous expressions that with our
assumptions there is no contribution to Eq.(\ref{cvariance}) coming from the
non-universal short periodic orbits.

The classical parameter $\alpha$ in Eq.(\ref{cvariance}) is generally, like
$\bar{\rho}$, a function of $X$ and of the energy. It depends on the system
and the particular parametric variation under study. For example, for a
two-dimensional billiard of area ${\cal A}$ and perimeter ${\cal L}$,
consider a general deformation which moves the boundary (parametrized by
$s$, $0\le s < {\cal L}$) by an amount $g(s) dX$ in the normal direction.
The quantities $Q_p$ for periodic orbits with period $T$ are given
in this case by
\begin{equation}
Q_p = - \frac{T {\cal L} \hbar^2 k^2}{2 m {\cal A}} \langle g \rangle
      + 2 \hbar k \sum_{i=1}^{n_p} \cos \theta_{i} \, g(s_i)
\end{equation}
where $\langle f \rangle = (1/{\cal L}) \int_0^{\cal L} f(s) d s$,
$k$ is the wave number corresponding to $\epsilon$, $n_p$ is the
number of bounces of the orbit, and $s_i$ and $\theta_{i}$ denote
the points of reflection and the angle of the trajectory with
the tangent to the boundary at these points, respectively.
After calculating $\langle Q_p^2 \rangle $ from this expression
we obtain the following result for $\alpha$
\begin{equation} \label{alpha}
\alpha = \frac{4 {\cal L} (2 m E)^{3/2}}{\pi m {\cal A}} \left( {\cal C}_1
\langle g^2 \rangle  -
\ {\cal C}_2 \langle g \rangle ^2 \right) \ .
\end{equation}
The constants ${\cal C}_1$ and ${\cal C}_2$ contain dynamical information about
the periodic orbits,
\begin{eqnarray}
{\cal C}_1 & = & \frac{2}{3} \left[ 1 + 2 \sum_{\tau=1}^{\infty} f(\tau)
\right] \\
{\cal C}_2 & = & \frac{\pi^2}{16} \left[ 1 + 2 \sum_{\tau=1}^{\infty} f(\tau)
- \frac{\sigma_{n_p}^2}{\overline{n_p}} \right] \ ,
\end{eqnarray}
with
\begin{equation}
f(\tau) = \frac{\langle g(s_i) \cos \theta_i \, g(s_{i+\tau}) \cos \theta_{i+\tau}
\rangle - (\pi^2 /16) \langle g \rangle^2 }
{(2/3) \langle g^2 \rangle  - (\pi^2 / 16) \langle g \rangle^2 } \ .
\end{equation}
The numerical factors in these expressions come from the ergodic phase-space
averages $\langle \cos \theta \rangle^2 = \pi^2 /16$ and
$\langle \cos^2 \theta \rangle = 2/3$. By definition, $f(0)=1$ and
$f(\tau) \rightarrow 0$ for large $\tau$. The quantities $\sigma_{n_p}^2$ and
$\overline{n_p}$ are
the variance and the average of the number of bounces of periodic orbits with
period $T\rightarrow \infty$, respectively. In the extreme case where the
correlations between the consecutive segments of an orbit are neglected
(the uncorrelated, ``random'' case), 
then ${\cal C}_1 = 2/3$ and ${\cal C}_2 = 0$.

Notice from Eqs.(\ref{cvariance}) and (\ref{alpha}) the energy dependence
$E^{3/2}$ of the average variance of the velocity \cite{blm} (to leading order 
${\bar \rho} = {\cal A}/4 \pi$ for two-dimensional billiards in 
dimensionless units $\hbar=2 m = 1$). This dependence holds for
arbitrary deformations of chaotic two-dimensional billiards, but other
dynamical properties of a system or other parametric variations may produce
different energy dependences (for example, for integrable billiard systems
we obtain $\langle v_X^2 \rangle \propto E^2$, while for chaotic billiards
with Aharonov-Bohm fluxes $\langle v_X^2 \rangle
\propto \sqrt{E}$
\cite{blm}). Fig.2 shows the (normalized) variance of the velocity for a
chaotic billiard in comparison with the semiclassical approximation
(\ref{alpha}) with ${\cal C}_1 = 2/3$ and ${\cal C}_2 = 0$.
A linear increase is observed, in agreement with the predicted 
energy dependence. Moreover, the slope agrees with the simple
semiclassical estimate. The variation with energy also implies that in
general it is not possible to simultaneously set to one the mean spacing and
$\langle v_X^2 \rangle$ for arbitrary energies. All parametric correlators
should therefore be computed in an energy window.

In conclusion, we have introduced a universal invariant way of scaling the
parameter-dependent correlators in quantum mechanics. We have moreover
obtained expressions which allow for an explicit implementation of this scaling
procedure and predicted a universal energy dependence of the variance of the
velocities for deformations of a chaotic billiard. Contrary to the density
of states which depends only on global geometrical properties of phase
space, the derivation of Eq.(\ref{cvariance}) assumes a (fully) chaotic
dynamics. Moreover the parametric-invariant scaling procedure is unique only
in the situation of generic chaotic and weakly disordered systems with a
Gaussian distribution of the velocities. Our results, in particular
the predictions concerning two-dimensional billiards, may be tested
experimentally in quantum dots, metallic grains or microwave cavities.

We would like to thank A.\ B\"acker, O.\ Bohigas and M.\ Robnik for useful
and stimulating discussions. We moreover thank A.\ B\"acker for the spectral
data of the Robnik-lima\c{c}on billiard, and H.\ Schanz for the spectral
data of the Sinai billiard. M. S. acknowledges financial support by the
Alexander von Humboldt-Stiftung and the Deutsche Forschungsgemeinschaft.

\begin{figure}
\begin{center}
\leavevmode
\epsfysize=3.8in
\epsfbox{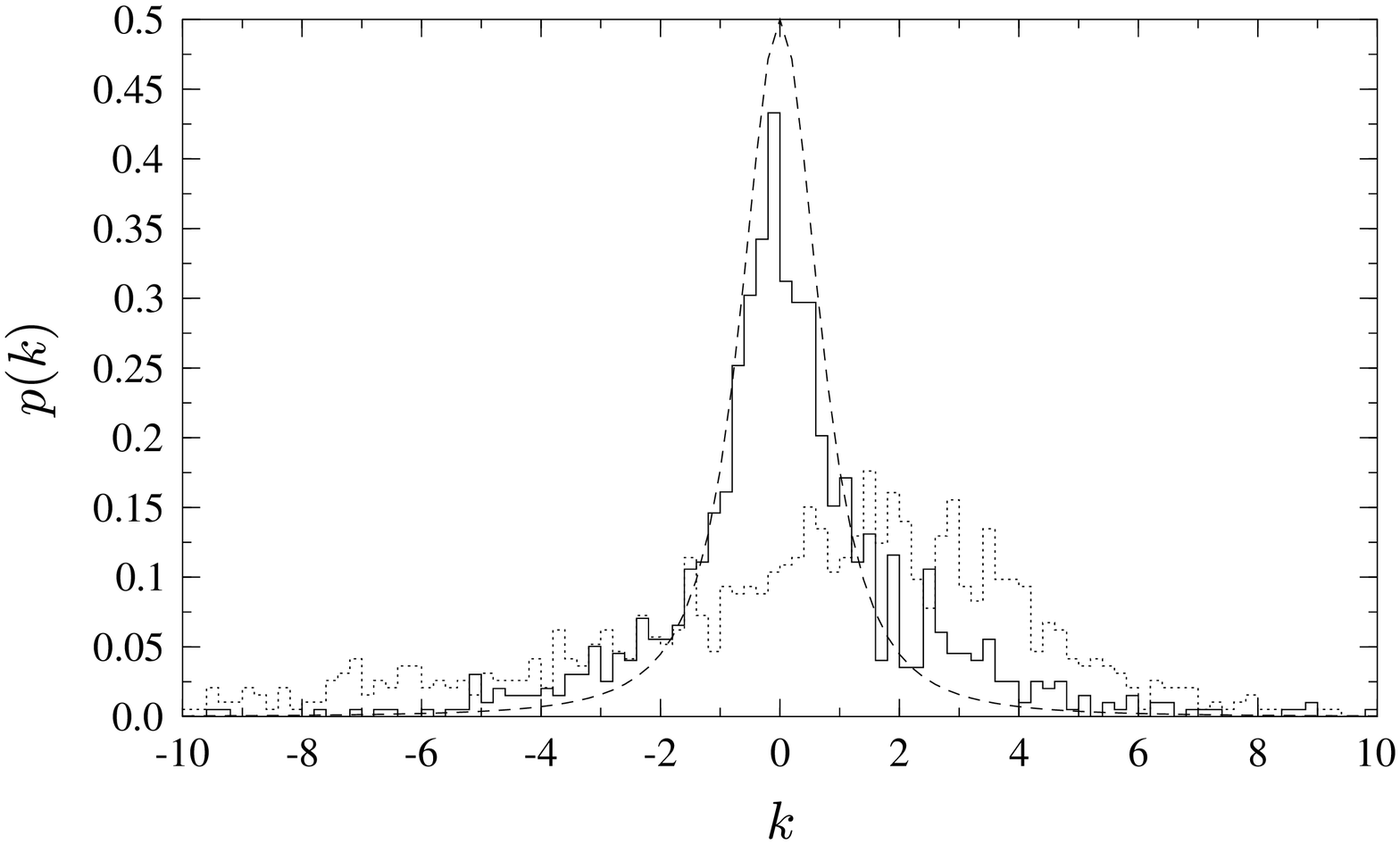}
\end{center}
\vspace{2.5cm}
\caption{Curvature distributions from the eigenvalues $n=2001$ to $n=3000$
  of the odd symmetry class of the Robnik-lima\c{c}on billiard at $X=\lambda =
  0.49$; dotted histogram: $k_x$ (non-invariant), continuous histogram:
  $k_\mu$ (invariant). Dashed curve: RMT prediction Eq.(\ref{crmt}).}
\label{fig1}
\end{figure}

\newpage

\begin{figure}
\begin{center}
\leavevmode
\epsfysize=3.8in
\epsfbox{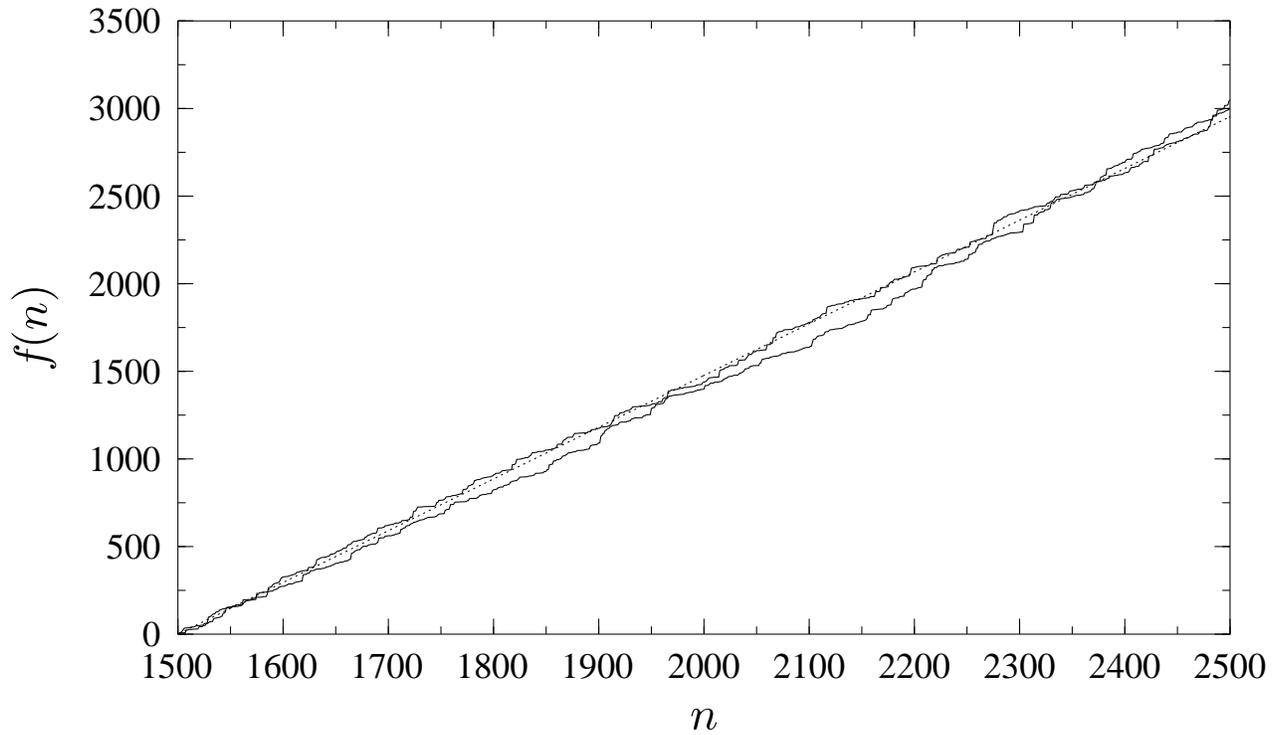}
\end{center}
\vspace{2.5cm}
\caption{The quantity $f(n) = \sum_{j=n_0}^n
(\partial \epsilon_j /\partial r)^2 
  / \epsilon_j^{3/2}$, in dimensionless units, for the eigenvalues of
the two symmetry classes of a quarter Sinai billiard at radius
$r=0.5$ and $n_0 =1500$ (full lines) in comparison with the 
semiclassical approximation (dotted line).}
\label{fig2}
\end{figure}

\end{document}